\documentclass[epj-spec]{svjour}
\usepackage{graphics}
\newcommand{\be}{\begin{equation}}
\newcommand{\ee}{\end{equation}}
\newcommand{\bea}{\begin{eqnarray}}
\newcommand{\ena}{\end{eqnarray}}

\newcommand{\pT}{{p_{_T} }}
\newcommand{\dsig}{\frac{d \sigma}{d {\vec p}_{_T} d \eta}}
\newcommand{\dsigd}{\frac{d \sigma^{\hbox{(dir)}}}{{d {\vec p}_{_T} d \eta}}}
\newcommand{\dsigf}{\frac{d \sigma^{\hbox{(frag)}}}{{d {\vec p}_{_T} d \eta}}}
\newcommand{\dsigij}{\frac{d {\widehat \sigma}_{ij}}{{d {\vec p}_{_T} d \eta}}}
\newcommand{\dsigijk}{\frac{d {\widehat \sigma}_{ij}^{k}}
                           {d {\vec p}_{_T} d \eta}}
\newcommand{\kd}{K^{\hbox{(dir)}}}
\newcommand{\kf}{K^{\hbox{(frag)}}}
\newcommand{\MSB}{{\overline {MS}}}
\newcommand{\LMS}{\Lambda_{_{_\MSB}}}

\newcommand{\alfspi}{\frac{\alpha_s(\mu_{R})}{2 \pi}}

\begin{document}
\title{Investigating the high energy QCD approaches for prompt photon  production at the LHC}
\author{M.V.T.
Machado\inst{1}\fnmsep\thanks{\email{magno.machado@unipampa.edu.br}} \and
C. Brenner Mariotto\inst{2}\fnmsep\thanks{\email{mariotto@fisica.furg.br}} }
\institute{Centro de Ci\^encias Exatas e Tecnol\'ogicas, Universidade Federal do Pampa.
Campus de Bag\'e, Rua Carlos Barbosa. CEP 96400-970. Bag\'e, RS, Brazil \and
Departamento de F\'\i sica, Universidade Federal do Rio Grande. Box 474, Rio Grande, RS CEP 96201-900, Brazil}
\abstract{
We investigate the rapidity and transverse momentum distributions of  the prompt
photon production at the CERN LHC energies considering the current perturbative QCD
approaches for this scattering process.  Namely, we compare the predictions from  the
usual NLO pQCD calculations to the the color dipole formalism, using distinct dipole cross
sections. Special attention is paid to parton
saturation models at high energies, which are expected to be important at the forward rapidities in $pp$ collisions ($\sqrt{s}=14$ TeV) at the LHC.
} 
\maketitle
\section{Introduction}
\label{intro}

There are recent experiments collecting {\em prompt} photon data, as the PHENIX collaboration at $\sqrt{s} = 200$~GeV at RHIC (collecting data both for the inclusive~\cite{Adler:2005} and the isolated
case~\cite{Okada:2005})  and  the D$\emptyset$  collaboration~\cite{Abazov:2005wc} ($\sqrt{s} = 1.96$ GeV) in the Tevatron (measuring isolated prompt photons whose
transverse momenta $p_{T}$ range from 23 to about 300 GeV). A photon is said to be {\em prompt} if
it does not originate from the decay of a hadron, such as a
$\pi^0$ or $\eta$, itself produced with large transverse momentum. From theoretical point of view, the cross sections for producing such isolated photons have been proven to still fulfill the factorisation property, and are finite to all orders in perturbation theory. The study of prompt photons carrying large values of transverse momentum has long history and it is related to deep-inelastic-lepton scattering (DIS), Drell-Yan process production and jet production as an important probe of
short-distance hadron dynamics. The great appeal of prompt photons is that they are point-like,
colorless probes of the dynamics of quarks and gluons, once that escape unscathed through
the colored medium of the high-energy collision. Thus, they can be a powerful probe
of the initial state of matter created in heavy ion collisions,
since they interact with the medium only electromagnetically and
therefore provide a baseline for the interpretation of jet-quenching
models.
Prompt photons offer access to the spin-dependent and spin-averaged gluon densities of hadrons, where one of the two leading-order partonic direct subprocesses feeds directly from the gluon parton density through the `Compton' subprocess,
$q g \rightarrow \gamma q$. It should be stressed that the simplicity of prompt photons is
compromised with the fragmentation contributions, where fragmentation is a long-distance process in which a hard photon brems off a final-state quark (or gluon).  The photon emerges as part of a jet if the
opening angle between the quark and photon is too small and this `showering' process
is parameterized by a non-perturbative single-photon fragmentation
function $D_{\gamma}(z, \mu_F)$.

Calculations exist of prompt single photon production at next-to-leading order (NLO)
in QCD for both the direct and the fragmentation contributions, as the approach is encoded in the Monte Carlo program JETPHOX~\cite{Aurenche:2006vj,Catani:2002ny,cocorico}. NLO pQCD calculation provides a fairly good description of the world-data on $p_T> 5$~GeV hadroproduction spectra~\cite{Aurenche:2006vj}, where the mid-rapidity cross section is sensitive to parton distributions at  $x\sim 2\,p_T/\sqrt{s}$ and $Q^2\sim p_T^2$ and at a given $Q^2$, other values of $x$ can be probed by measuring photons at non-zero rapidity. Accordingly, the gluon densities in a proton and in a nucleus are fundamental ingredients in order to compute the corresponding hard-process observables in $pp$, $pA$ and $AA$ collisions. The gluon distribution $xg(x,Q^2)$ is fairly well known within a few percent accuracy in the range $x\sim 10^{-5}$--$10^{-2}$ and $Q^2\sim 10$--$10^5$~GeV$^2$~\cite{Pumplin:2002vw}, which is precisely the kinematical domain covered by most hard processes at the LHC. However, very little is known on the nuclear gluon density per nucleon, $xg_A(x,Q^2)$~\cite{Armesto:2006ph}. 
One way to better constrain the nuclear gluon pdf is though prompt photon production. 
In Ref. \cite{Arleo} the prompt photon production in $p$--Pb collisions at LHC
and $d$--Au collisions at RHIC  has been explored, using that the nuclear
production ratio of isolated photons in $p$--A collisions can be simply
approximated as a linear combination of gluon pdf's and structure functions in
the nucleus A over those in a proton. Performing the calculation in pQCD at NLO
they have checked that such an approximation is correct up to a few-percent
accuracy and makes this observable an ideal tool to measure gluon shadowing at
RHIC and at LHC. A LO QCD study of prompt photon production in pA and AA
collisions including more recent nuclear pdf's is done in ref. \cite{BMG:2008},
and it is shown that the amount of shadowing and antishadowing is very distinct
for the various nuclear pdf's available, so these processes are usefull to
discriminate among the different nuclear pdf's. Moreover, a precise knowledge of
the gluon pdf's is crucial for a better understanding of the QCD evolution, mostly the possible non-linear QCD evolution at small $x$ and small $Q^2$ , where the gluon density in the nucleus becomes large and starts to saturate at momentum scales near $Q_{sat}$ (the so-called ``saturation scale'')~\cite{Iancu:2003xm}. 
The influence of nonlinear gluon evolution in prompt photon production is estimated in Ref. \cite{BMGprc2007}, where we find an enhancement of the prompt photon cross section for LHC energies.

Recently, it has been verified that the color dipole approach can successfully describe inclusive photon
production in $pp$ collisions at midrapidities. In Ref. \cite{Kope_gamma} it was showed that both direct photon production and Drell-Yan dilepton pair production processes can be described within the same
color dipole approach without any free parameters. Such a formalism, developed in \cite{bb} for the case of the total and diffractive cross sections, can be also applied to
radiation \cite{hir}. Although in the process of electromagnetic
bremsstrahlung by a quark no real quark dipole participates, the cross section
can be expressed via the more elementary cross section
$\sigma_{dip}$ of interaction of a $Q\bar{Q}$ dipole. It was also showed
that the color dipole formulation coupled to the DGLAP evolution
provides a better description of data at large transverse momentum
compared to simple saturation dipole models. In contrast to the parton model, any photon
fragmentation function was included for computing the cross
section, since the dipole formulation already incorporates all
perturbative (via Pomeron exchange) and non-perturbative radiation
contributions. Using similar approach, i.e. the Color Glass Condensate (CGC) formalism, in Ref. \cite{VM} the authors presented the predictions for prompt photon production using a model for the scattering dipole amplitude which describes quite well the hadron production. As stressed in Refs. \cite{jamal_photon,betemps,jamal_tam}, electromagnetic probes of the CGC are crucial to determine the dominant physics in the forward region at RHIC and LHC.  They estimated the ratio $R_{hA}$ for photon production at forward rapidities for RHIC and LHC energies and compare its behavior with that predicted for hadrons and predicted the photon to pion production ratio and study its $p_T$ dependence. Therefore, such an approach is quite promissing to predict observables to measured at the LHC.

Our goal in this contribution is twofold. Firstly, we perform a comparison between the NLO QCD approach to the color dipole formalism pointing out the possible deviations and their origin. Secondly, we investigate in detail the color dipole approach using distinct implementations for the color dipole cross section, discussing several phenomenological aspects.  The paper is organized as follows. In next section we summarize the main formula for both NLO pQCD approach and color dipole formalism. In last section we show our numerical results and predictions.  Finally, we summarize our main conclusions.

\section{Prompt photon production}

Let us summarize the theoretical treatment for prompt photon production in $pp$ collisions in high energy colliders. We start with the usual parton model description of underlying processes and after that we present the alternative color dipole formalism. We call attention that the dipole approach is well suited for high-energy processes, i.e. small $x_{2}$, and its range of validity is expected to be near $x_{2}<0.1$. The main advantage in such an approach is to describe simultaneously the direct photon and dilepton production in the same framework.

\subsection{NLO QCD approach}
Schematically, the production of a prompt photon proceeds via two
mechanisms. The first one is the `direct' mechanism, where the photon
behaves as a high $p_{T}$ colourless parton, i.e. it takes part in the hard
subprocess, and it is most likely to be well separated from any hadronic
environment. The second one is called `fragmentation' mechanism, where now the
photon behaves as a kind of (anomalous) hadron, i.e. it results from the
collinear fragmentation of a coloured high $p_{T}$ parton, and is it most
probably accompanied by hadrons - unless the photon carries away most of the
transverse momentum of the fragmenting parton. The fragmentation contribution emerges from the calculation of the higher order corrections to direct mechanism in the perturbative expansion in powers of the strong
coupling $\alpha_{s}$. At higher orders, final state multiple collinear
singularities appear in any subprocess where a high $p_{T}$ parton of species
$k$ (quark or gluon) undergoes a cascade of successive collinear splittings
ending up with a splitting into a photon. These singularities are factorised to
all orders in $\alpha_s$ according to the factorisation theorem, and absorbed
into fragmentation functions of parton $k$ to a photon,
$D_{\gamma/k}(z,M_{_F})$, defined in some arbitrary fragmentation scheme, at
some arbitrary fragmentation scale $M_{_F}$. The point-like coupling of the
photon to quarks is responsible for the well-known anomalous behaviour of
$D_{\gamma/k}(z, M_{_F})$, roughly as $\alpha_{em}.\alpha_{s}^{-1}(M_{_F})$
when the fragmentation scale $M_{_F}$, chosen of the order of a hard scale of
the subprocess, is large compared to ${\cal O}(1$~GeV). The differential cross
section in transverse momentum $\pT$ and rapidity $\eta$ can thus be written as:
\begin{equation}
\sigma(pp\rightarrow \gamma+X)
=
\sigma^{(\mathrm{dir})}(\mu_{R},M,M_{_F})
+\sum_{k=q,\bar{q},g}\sigma^{(\mathrm{frag})}_{k}(\mu_{R},M,M_{_F})
\otimes D_{\gamma/k}(M_{_F})
\label{eq1}
\end{equation}
where $\sigma^{(\mathrm{frag})}_{k}$ describes the production of a parton $k$ in a
hard collision. The arbitrary parameters $\mu_{R}$, $M$ and $M_{_F}$
are respectively the renormalisation, initial-state factorisation, and
fragmentation scales. The dependence of the NLO predictions with respect
to $\mu_{R}$, $M$ and $M_{_F}$ is simplified if we take all scales to be equal and they will be noted $\mu$. We rely on the calculation of both direct and
fragmentation at next-to-leading order (NLO) accuracy \cite{Aurenche:1998gv,cocorico}, which takes the
form ($\eta$ is the photon rapidity)
\be
\dsig = \dsigd\ + \ \dsigf
\label{eq:sig}
\ee
where
\bea
\dsigd
 &=& \sum_{i,j=q,\bar{q},g} \int dx_{1} dx_{2}
\ F_{i/h_1}(x_{1},M)\ F_{j/h_2}(x_{2},M)
\alfspi \nonumber \\
&\times &\left[ \dsigij + \alfspi \kd_{ij} (\mu_{R},M,M_{_F}) \right]
\label{eq:dir}
\ena
and
\bea
\dsigf &=& \sum_{i,j,k=q,\bar{q},g} \int dx_{1} dx_{2}\frac{dz}{z^2}
\ F_{i/h_1}(x_{1},M)\ F_{j/h_2}(x_{2},M)\ D_{\gamma/k}(z, M_{_F})
\left( \alfspi \right)^2\nonumber \\
 &\times & \left[ \dsigijk \ +
\ \alfspi \kf_{ij,k} (\mu_{R},M,M_{_F}) \right]
\label{eq:brem}
\ena
where $F_{i/h_{1,2}}(x,M)$ are the parton distribution functions of  parton
species $i$ inside the incoming hadrons $h_{1,2}$, at momentum fraction $x$ and
factorisation scale $M$; $\alpha_{s}(\mu_{R})$ is the strong coupling  defined
in the $\overline{\mbox{MS}}$ renormalisation scheme at the  renomalisation
scale $\mu_{R}$.  The knowledge of  $\LMS$, e.g. from deep-inelastic scattering
experiments, completely specifies  the NLO expression of the running coupling
$\alpha_{s}(\mu_{R})$. The NLO correction terms to direct and fragmentation,
$\kd_{ij}$~\cite{ABDFS} 
 and $\kf_{ij,k}$~\cite{ACGG} respectively, are known
and their expressions in the
$\overline{\mbox{MS}}$ scheme will be used. In the calculation of numerical results in next section, we will show the direct and fragmentation contribution, presenting the uncertainties associated with different choices for the scale $\mu$.

\subsection{Color dipole formalism}

The color dipole formalism can be used to compute direct photon production as it is also applied to
describe radiation processes \cite{hir}. The transverse momentum $p_{T}$ distribution of photon
bremsstrahlung in quark-nucleon interactions, integrated over the
final quark transverse momentum, was derived in \cite{p6} in terms
of the dipole formalism,
 \begin{eqnarray}
\frac{d \sigma^{qN}(q\to q\gamma)}{d(ln \alpha)\,d^{2}\vec{p}_{T}}&=&\frac{1}{(2\pi)^{2}}
\sum_{in,f}\sum_{L,T}
\int d^{2}\vec{r}_{1}d^{2}\vec{r}_{2}e^{i \vec{p}_{T}.(\vec{r}_{1}-\vec{r}_{2})}\phi^{\star T,L}_{\gamma q}(\alpha, \vec{r}_{1})
\phi^{T,L}_{\gamma q}(\alpha, \vec{r}_{2}) \nonumber \\
&\times &\frac{1}{2}\{
\sigma_{dip}(x,\alpha r_{1})+\sigma_{dip}(x,\alpha r_{2})\}-\frac{1}{2}\sigma_{dip}(x,\alpha(\vec{r}_{1}-\vec{r}_{2})), \label{di}
\end{eqnarray}
where $\vec{r}_{1}$ and $\vec{r}_{2}$ are the quark-photon transverse
 separations in the two radiation amplitudes contributing to the cross
 section, $\sigma_{dip}$. The parameter $\alpha$ is the
 relative fraction of the quark momentum carried by the photon, and is
 the same in both amplitudes, since the interaction does not change
 the sharing of longitudinal momentum.  In the equation above, $T$
 stands for transverse and $L$ for longitudinal photons. The energy
 dependence of the dipole cross section, which comes via the variable
 $x=2 (p_1\cdot q)/s=(p_T/\sqrt{s})\,e^{-y}$, where $p_1$ is the projectile four-momentum and $ q$ is
the four-momentum of the dilepton, is generated by additional
radiation of gluons which can be resummed in the leading $\ln(1/x)$
approximation.

In Eq.~(\ref{di}) the light-cone wavefunction of the projectile
quark $\gamma q$ fluctuation has been decomposed into  transverse
$\phi^{T}_{\gamma q}(\alpha, \vec{r})$ and longitudinal
$\phi^{L}_{\gamma q}(\alpha, \vec{r})$ components,  and an average
over the initial quark polarization and sum over all final
polarization states of quark and photon is performed. For direct photons, only transverse component contributes and this wavefunction component $\phi^{T}_{\gamma q}(\alpha, \vec{r})$ can
be represented at the lowest order as:
\begin{eqnarray}
\sum_{in,f}\phi^{T\star}_{\gamma q}(\alpha, \vec{r}_{1})\phi^{T}_{\gamma q}(\alpha, \vec{r}_{2})
= \frac{\alpha_{em}}{2\pi^{2}} \,\left \{ m^2_{q}\alpha^{4}K_{0}(\epsilon r_{1})K_{0}(\epsilon r_{2})+[1+(1-\alpha)^{2}]\epsilon^{2}\frac{\vec{r}_{1}.\vec{r}_{2}}{r_{1}r_{2}}
K_{1}(\epsilon r_{1})K_{1}(\epsilon r_{2})\right \}, \nonumber
\label{wave}
\end{eqnarray}
in terms of transverse separation $\vec{r}$ between photon $\gamma $
and quark $ q$ and the relative fraction $\alpha$ of the quark
momentum carried by the photon.  Here $K_{0,1}(x)$ denotes the
modified Bessel function of the second kind and it has been introduced
the auxiliary variable $\epsilon^{2}=\alpha^{2}
m_{q}^{2}$, where $m_{q}$ is an effective quark mass. Here, we take $m_{q}=0.2$ GeV for direct
photon production.

The hadron cross section can be obtained from the elementary
partonic cross section Eq.~(\ref{di}) summing up the
contributions from quarks and antiquarks weighted with the
corresponding parton distribution functions (PDFs),
\begin{eqnarray}
\frac{d \sigma (pp\to \gamma X)}{dyd^{2}\vec{p}_{T}}= \int_{x_{1}}^{1}\frac{d\alpha}{\alpha} F_{2}^{p}(\frac{x_{1}}{\alpha},\mu^2)\,
\frac{d \sigma^{qN}(q\to q\gamma)}{d(ln \alpha)\,d^{2}\vec{p}_{T}},\
\label{con}
\end{eqnarray}
where the PDFs of the projectile have entered in a combination which can be
written in terms of proton structure function $F_{2}^{p}$. For the hard scale $\mu$ entering in the proton
structure function in Eq.~({\ref{con}}), we take $\mu^{2}=p^{2}_{T}$ and the energy scale
$x$ of the dipole cross section entered in Eq.~(\ref{di}) was set $x=x_{2}$. For the proton structure function in Eqs.~(\ref{con}) we have taken the recent ALLM parametrization \cite{ALLMnew}, which is valid in the kinematic range we are interested in. The sensitivity to a different choice for $F_2$ is very small.

A crucial ingredient in the color dipole calculations is the dipole cross section. It is theoretically unknown, although several parametrizations have been proposed. For our
purposes, here we consider some analytical parametrizations. The simplest parametrizations rely on the geometric scaling property. That is, the dipole cross section is a function of a scaling variable $rQ_{sat}(x)$ where $Q_{sat}$ is the so called saturation scale. It defines the transverse momentum scale where parton recombination physics is relevant and in general is modeled as $Q_{sat}\propto x^{-\lambda/2}$ (it grows with energy). A common feature on these models is that for decreasing $x$, the
dipole cross section saturates for smaller dipole sizes, and that at
small $r$, as perturbative QCD implies, $\sigma\sim r^{2}$ vanishes, i.e. the color transparency phenomenon \cite{bb}. In a general form, the dipole cross section can be parametrized as,
\begin{equation}
\sigma_{dip}(x,\vec{r};\gamma)=\sigma_{0}\left[ 1-\exp\left(-\frac{r^{2}Q_{sat}^{2}}{4}\right)^{\gamma_{\mathrm{eff}}}\,\right], \hspace{1cm} Q_{sat}^2(x)=\left(\frac{x_0}{x}\right)^{\lambda},
\label{gbw}
\end{equation}
where the quantity $\gamma_{\mathrm{eff}}$ is the so-called effective anomalous dimension. For the celebrated GBW parametrization \cite{gbw}, $\gamma_{\mathrm{eff}} = 1$ and the remaining parameters are fitted to DIS HERA data at small $x$. This
parametrization gives a quite good description of DIS data at $x<10^{-2}$.
A very recent adjust \cite{MKW} using GBW model for $0.25<Q^2<45$  GeV$^2$ and using quark masses $m_q=0.14$ GeV and $m_c=1.4$ GeV gives $\sigma_0= 23.9$ mb,  $x_0=1.11\times 10^{-4}$, $\lambda = 0.287$ ($\chi^2/\mathrm{dof}=1.58$).

The main difference among the distinct phenomenological models using parametrizations as Eq. (\ref{gbw}) comes from the  predicted behavior for the anomalous dimension, which determines  the  transition from the non-linear to the extended geometric scaling regimes, as well as from the extended geometric scaling to the DGLAP regime. It is  the behavior of $\gamma$ that determines the fall off with increasing $p_T$ of the hadronic cross section. The current models in the literature consider the general form $\gamma_{\mathrm{eff}} = \gamma_{sat} + \Delta (x,r;p_T)$, where $\gamma_{sat}$ is the anomalous dimension at the saturation scale and $\Delta$ mimics the onset of the geometric scaling region and DGLAP regime. One of the basic differences between these models is associated to the behavior predicted for $\Delta $. While the models proposed in Refs. \cite{iim,kkt,dhj} assume that $\Delta$ depends on terms which violate the geometric scaling, i.e. depends separately on $r$ and rapidity $Y=ln(1/x)$,  the model recently proposed in Ref. \cite{buw} by D. Boer et al. consider that it is a function of $r \,Q_{sat}$.
\begin{eqnarray}
\label{BUWeq}
\gamma_{\mathrm{eff}}= \gamma_{sat}+(1-\gamma_{sat})\frac{(\omega^a-1)}{(\omega^a-1)+b},
\end{eqnarray}
where $\omega  \equiv p_T/Q_{sat}$ and the two free parameters $a=2.82$ and $b=168$ are fitted in order do describe the RHIC data on hadron production. The remaining parameters  are taken from the GBW parametrization. Thus, for large $p_T$ the Boer at al. dipole cross section reproduces the GBW model. They differ at small $p_T\leq Q_{sat}$ compared to the saturation scale, where $\gamma_{\mathrm{eff}}\approx \gamma_{sat}\simeq 0.63$.

In order to investigate the effect of QCD evolution in the dipole cross section we also will include in our studies the impact parameter saturation model \cite{MKW}. It has been successful in describing the exclusive vector meson production. The dipole cross section in this model is given by:
\begin{equation} \label{eq:dsigmad2b}
\sigma_{dip} (x,r) = 2\int d^2b\,\left[1-\exp\left(-\frac{\pi^2}{2N_c}r^2\alpha_S(\mu^2)xg(x,\mu^2)T(b)\right)\right],
\end{equation}
where the scale $\mu^2$ is related to the dipole size $r$ by $\mu^2=4/r^2+\mu_0^2$.  The gluon density, $xg(x,\mu^2)$, is evolved from a scale $\mu_0^2$ up to $\mu^2$ using LO DGLAP evolution without quarks.
The initial gluon density at the scale $\mu_0^2$ is taken in the form $xg(x,\mu_0^2) = A_g\,x^{-\lambda_g}\,(1-x)^{5.6}$. The values of the parameters $\mu_0^2$, $A_g$, and $\lambda_g$ are determined from a fit to $F_2$ data.  For the light quarks, the gluon density is evaluated at $x=x_B$ (Bjorken-$x$), while for charm quarks, $x=x_B(1+4m_c^2/Q^2)$. The LO formula for the running strong coupling $\alpha_S(\mu^2)$ is used, with three fixed flavours and $\Lambda_{\mathrm{QCD}}=0.2$ GeV.
The proton shape function $T(b)$ is normalised so that $\int d^2b\,T(b) = 1$ and a Gaussian form for $T(b)$ is considered, $T_G(b) = \frac{1}{2\pi B_G}\mathrm{e}^{-\frac{b^2}{2B_G}}$.
 A recent adjust using b-SAT \cite{MKW} model for $0.25<Q^2<650$  GeV$^2$ using quark masses $m_q=0.14$ GeV and $m_c=1.4$ GeV gives $\mu_0^2=1.17$ GeV$^2$,  $A_g=2.55$  and $\lambda_g=0.020$,  producing a good quality ajust with  $\chi^2/\mathrm{dof}=1.21$. In next section we will use the phenomenological models presented here to compute the $p_T$ spectrum of direct photon production in $pp$ collisions.

\section{Numerical results}

\begin{figure}[t]
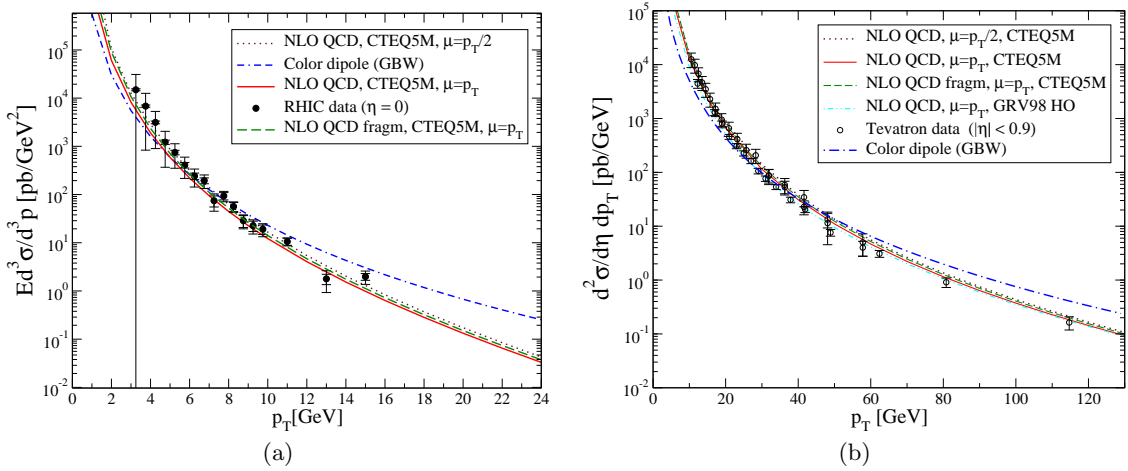

\begin{tabular}{cc}
\resizebox{0.5\columnwidth}{!}{\includegraphics{gammarhic_nlo_eta0_p.eps}} &
\resizebox{0.5\columnwidth}{!}{\includegraphics{gammatev_nlo_eta09_p.eps}} \\
(a) & (b)
\end{tabular}
\caption{(a) The invariant cross section for prompt photon production in $pp$ collisions at RHIC compared to NLO QCD and color dipole (using GBW model) results. (b) Predictions for the differential cross section $d^2\sigma(pp\rightarrow \gamma\,X)/dp_T\,d\eta$ using the NLO QCD and color dipole formalism at the Tevatron ($|\eta|<0.9$).}
\label{fig:1}
\end{figure}

Let us now present some numerical results concerning direct photon production in central rapidities. In Fig.~\ref{fig:1}-a, we show the result for the direct photon production in $pp$ collisions at RHIC obtained by the NLO QCD approach and color dipole picture. The invariant cross section, $Ed^3\sigma/d^3P$, is computed as a function of transverse momentum for central rapidity, $y=0$, at $\sqrt{s}=200$ GeV and data are taken from Ref. \cite{p3}. For the NLO QCD calculation, we test the theoretical uncertainty using two distinc scales: $\mu=p_T/2$ (dotted line) and $\mu=p_T$ (solid line). We have included also the photon fragmentation contribution \cite{Bourhis:1997yu} with scale $\mu=p_T$ (long dashed line). In these calculations, we used the CTEQ5M parton distribution functions (small changes using alternative PDFs sets). We verify relatively small uncertainties coming from set of PDFs and/or hard scale.  For the color dipole calculation, we reproduce the results on Ref. \cite{Kope_gamma} where the GBW dipole cross section is considered (dot-dashed line). The deviation between NLO QCD and color dipole starts to be sizable at transverse momenta of order 10 GeV, reaching one order of magnitude at $p_T=25$ GeV.
In Fig.~\ref{fig:1}-b, 
the differential cross section, $d^2\sigma/d\eta \,dp_T$, is computed for $pp$ collisions at the Tevatron, $\sqrt{s}=1.8$ TeV and $|\eta|<0.9$, for different choices of scale and parton distributions. The notation is the same as in the previous figure. The experimental points are CDF data \cite{p4}. The NLO QCD describes the data quite well, while the color dipole (GBW without evolution and with threshold correction) result seems to give an incorrect behavior. Next we show how this problem can be cured within the color dipole picture.

\begin{figure}[t]
\begin{tabular}{cc}
\resizebox{0.5\columnwidth}{!}{\includegraphics{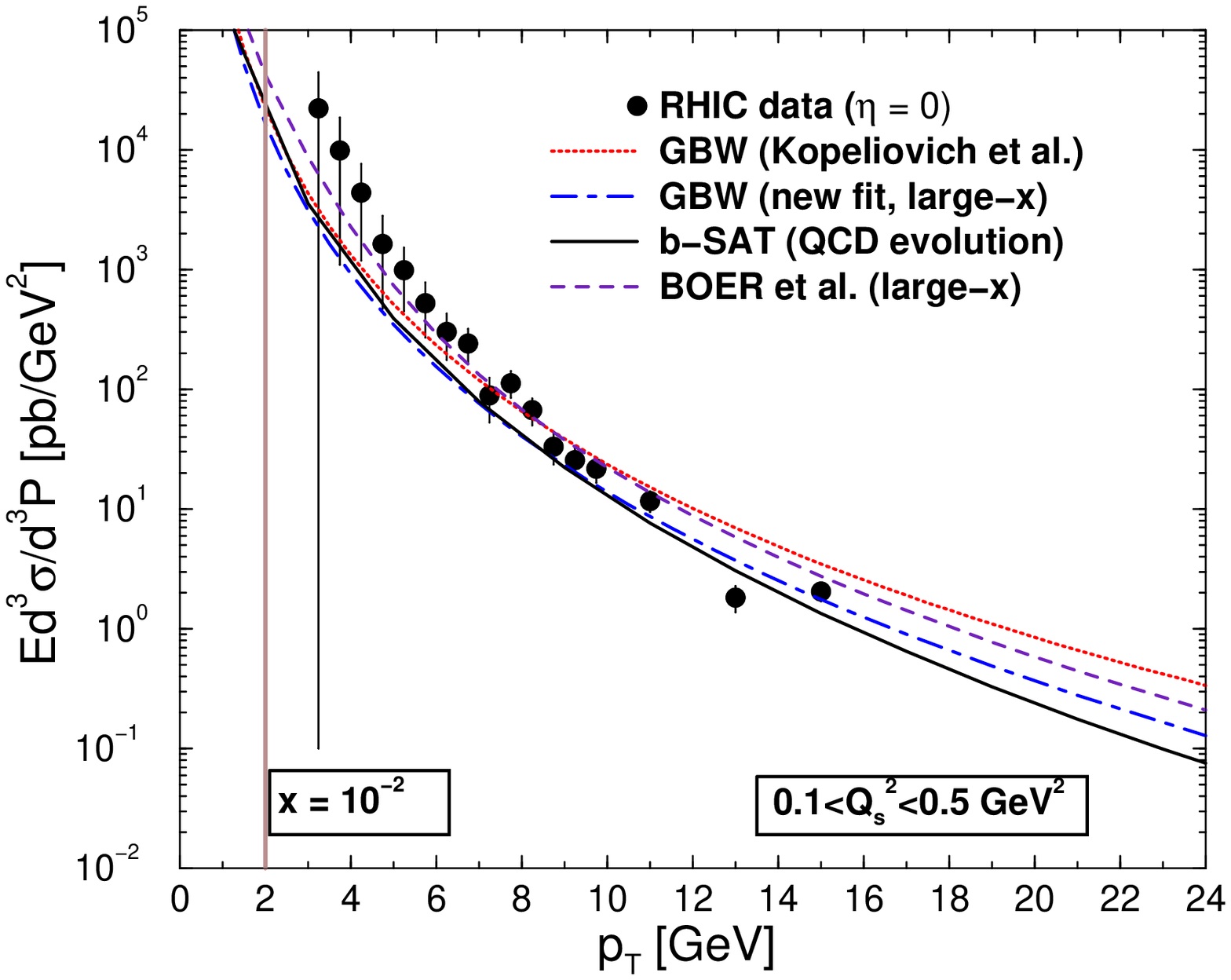}} & 
\resizebox{0.5\columnwidth}{!}{\includegraphics{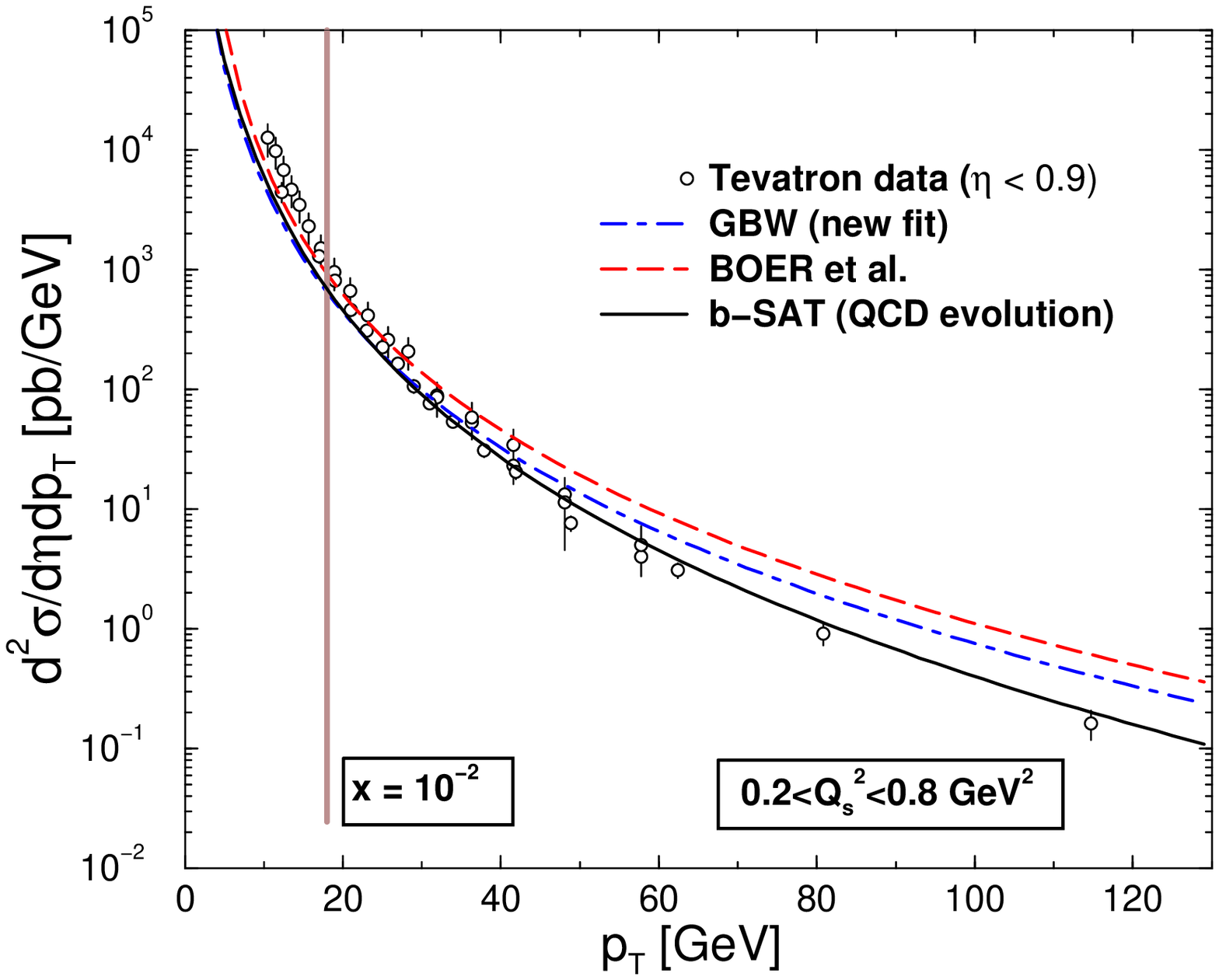}} \\
(a) & (b)
\end{tabular}
\caption{(a) The invariant cross section $Ed^3\sigma(pp\rightarrow \gamma\,X)/d^3P$ for $pp$ collisions at RHIC compared to different implementations of dipole cross sections within the color dipole picture (see text). (b) Predictions for the differential cross section $d^2\sigma(pp\rightarrow \gamma\,X)/d\eta\,dp_T$ using the several implementations for the dipole cross section for Tevatron energy.}
\label{fig:2}
\end{figure}

In Fig.~\ref{fig:2}-a we show the invariant cross section (central pseudorapidity $\eta =0$ at RHIC \cite{p3}, $\sqrt{s}=200$ GeV) obtained from several implementations of the dipole cross section taken from recent phenomenological works. For instance, we the following models: (i) the baseline calculation of Ref. \cite{Kope_gamma} (Kopeliovich et al., dotted solid line); (ii) the new fit with GBW dipole cross section, corrected by large-$x$ threshold (dot-dashed line); (iii) the impact parameter saturation model which includes QCD evolution in the dipole cross section, labeled here b-SAT (solid line); (iv) the D. Boer et al. model (labeled BOER et al., dashed line) which uses a running anomalous dimension.  The main deviation among the models occurs at large $p_T$. We first verify that the supposed incorrect behavior of GBW model discussed in Ref. \cite{Kope_gamma} is due to the lack of a large-$x$ threshold corrections. We have multiplied the GBW dipole cross section by a factor $(1-x_2)^5$ and this fact produces large suprression of cross section at large $p_T$. For sake of ilustration, we show the vertical line at $p_T\simeq 2$ GeV which corresponds to $x_2=p_T/\sqrt{s}=10^{-2}$, that is the expected limit of the color dipole approach. The RHIC data are all clearly within the large-$x$ region and, therefore, a threshold correction should be added in a consistent phenomenology. The BOER et al. model improves the data description al small $p_T$ but is similar to GBW towards large transverse momenta. The reason for that is that at large $p_T$ the effective anomalous dimension in both models coincide, giving $\gamma = 1$. The b-SAT model gives similar results as GBW at small $p_T$, wheras the data description is improved at large $p_T$. This property is already know in studies from Ref. \cite{Kope_gamma}, where the QCD evolution present in BGBK saturation model gave better data description. Finally, we show that saturation physics is not directly relevant for RHIC at midrapidity by writing down the values of the saturation scale in the kinematic range of data. We consider for simplicity the saturation scale from GBW mode, $Q_{sat}^2= (x_0/x_2)^{\lambda}=(x_0\sqrt{s}\,e^y/p_T)^{\lambda}$. We got $0.1\leq Q_{sat}^2\leq 0.5$ GeV$^2$, which is very small compared to the transverse momenta $4\leq p_T^2\leq 100$ GeV$^2$, Therefore, saturation effects do not play an important role at RHIC midrapidity.

In Fig.~\ref{fig:2}-b, we performed the same study for Tevatron. It is shown the differential cross section $d^2\sigma /d\eta dp_T$ as a function of transverse momentum for CDF energies $\sqrt{s}=1.8$ TeV  \cite{p4}. 
At lower transverse momentum  $p_{T}<40$ GeV the GBW
dipole model can reproduce rather fairly the experimental data, and at
higher $p_{T}$ values DGLAP present in the b-SAT evolution significantly improves the
results. It should be stressed that even at Tevatron energies, the large-$x$ effects are still important. We show the vertical line at $p_T\simeq 20$ GeV which corresponds to $x_2=10^{-2}$, telling us that the large $p_T$ data at Tevatron are out of small-$x$ region. Accordingly, we have used the dipole models corrected by threshold factor as referred before. The b-SAT model already contains such a correction by construction. The same statements about the role played by saturation efects remains valid for Tevatron at midrapidty, where the saturation scale is in the range $0.2\leq Q_{sat}^2\leq 0.8$ GeV$^2$ . As a partial conclusion, at low transverse momentum $p_{T}<10$ GeV saturation models predictions are
almost identical, but at higher $p_{T}$ the dipole parametrization
 including DGLAP evolution bends down towards the experimental
points improving the result.

\begin{figure}[t]
\begin{tabular}{cc}
\resizebox{0.5\columnwidth}{!}{\includegraphics{gammalhc_nlo_eta0_lowpt_dydpt_p.eps}} & 
\resizebox{0.5\columnwidth}{!}{\includegraphics{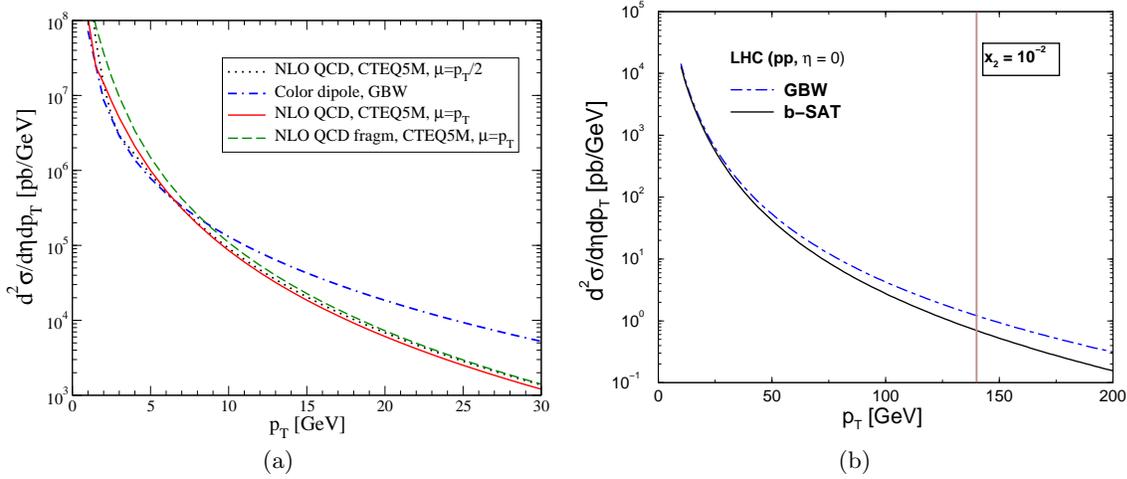}} \\
(a) & (b)
\end{tabular}
\caption{(a) Predictions for the differential cross section $d^2\sigma(pp\rightarrow \gamma\,X)/d\eta\,dp_T$ using the NLO QCD and color dipole formalism at the LHC. (b) The same 
using the two models for the dipole cross section for LHC energy (extended $p_T$ region).
}
\label{fig:3}
\end{figure}

In Fig.~\ref{fig:3}-a, the differential cross section, $d^2\sigma/d\eta \,dp_T$, is computed for $pp$ collisions at central rapidity at the LHC, $\sqrt{s}=14$ TeV, for the lower $p_T$ region up to $30$ GeV. 
The notation is the same as in figure \ref{fig:1}. It is verified that the sensivity on the scale for the NLO QCD calculation is larger than for RHIC and Tevatron, mostly at smaller $p_T$. The deviation between QCD and color dipole remains the same at the LHC. 
In Fig.~\ref{fig:3}-b we show the dipole models predictions (GBW and b-SAT) for
inclusive prompt-photon production at midrapidities, and for LHC
energies $\sqrt{s}=14$ TeV.  At lower transverse momentum  $p_{T}<30$ GeV the GBW
dipole model produces similar results as the b-SAT model and at
higher $p_{T}$ values DGLAP evolution becomes significantly important. It is verified that even at midrapidities at the LHC, the $x_2$ values are small as it can be shown by the vertical line at $p_T\simeq 130$ GeV. Once again, despite the saturation scale reaching values above a few GeV at central rapidities at the LHC, saturation effects should be negligible. The correct place to search such effects is looking at very forward rapitidies in the $pp$ mode or in the nuclear mode, where saturation scale should be enhanced by a factor $A^{1/3}$ (a factor 6 or 7 for Lead).

Finally, we will investigate semi-analytical calculations which are allowed in color dipole picture in the color transparency region. First, we can write the explicit expression for $p_T$ distribution using Eq. (\ref{con}) and the expressions for the transverse momentum $p_{T}$ distribution of photon
bremsstrahlung in quark-nucleon interactions, Eq. (\ref{di}), and transverse light cone wave function, Eq. (\ref{wave}). It reads as:
\begin{eqnarray}
\frac{d\sigma\,(pp\to \gamma
X)}{dy d^{2}\vec{p}_{T}} & = &
\frac{\alpha_{em}}{2\pi^2}\int_{x_{1}}^{1}\frac{d\alpha}{\alpha}
 F_{2}^{p}\left(\frac{x_{1}}{\alpha},Q^2=p_T^2\right) \nonumber\\
&\times & \left \{ m_q^2\alpha^4\left[\frac{{\cal I}_1}{(p_T^2+\varepsilon^2)}-\frac{ {\cal I}_2}{4\varepsilon} \right]
  +  [1+(1-\alpha)^2]\left[ \frac{\eta p_T \, {\cal I}_3}{(p_T^2+\varepsilon^2)} -\frac{{\cal I}_1}{2}+\frac{\varepsilon \,{\cal I}_2}{4}\right]
\right \},
\end{eqnarray}
 where we recall the auxiliary function $\varepsilon = \alpha \,m_q$. The quantities ${\cal I}_{1,2,3}$ are Hankel's integral transforms of order $0$ (${\cal I}_{1,2}$) and order $1$ (${\cal I}_{3}$) given by:
\begin{eqnarray}
{\cal I}_1 & = & \int_0^{\infty}dr\,rJ_0(p_T\,r)K_0(\varepsilon\,r)\, \sigma_{dip}(x_2,\alpha r),\hspace{1cm} {\cal I}_2  =  \int_0^{\infty}dr\,r^2J_0(p_T\,r)K_1(\varepsilon\,r)\, \sigma_{dip}(x_2,\alpha r),\nonumber \\
{\cal I}_3 & = & \int_0^{\infty}dr\,rJ_1(p_T\,r)K_1(\varepsilon\,r)\, \sigma_{dip}(x_2,\alpha r).
\label{ints}
\end{eqnarray}

If we consider that the current experimental results for midrapidities select a kinematic interval where the dipole cross section is dominated by the color transparency region, we can use the GBW parametrization and take its small-$r$ limit to compute analytically the integrals in Eq. (\ref{ints}).  In this case, we can take the approximation $\sigma_{dip}\approx \sigma_0(r^2\,Q_{sat}^2)$  in the region where $p_T\gg Q_{sat}$ and we get:
\begin{eqnarray}
{\cal I}_1  =  \sigma_0\,Q_{sat}^2\,\frac{(\varepsilon^2-p_T^2)}{(p_T^2+\varepsilon^2)^3}, \hspace{0.6cm}
 {\cal I}_2 =\sigma_0\,Q_{sat}^2\,\frac{4\varepsilon\,(\varepsilon^2-2p_T^2)}{(p_T^2+\varepsilon^2)^4},\hspace{0.6cm} {\cal I}_3  =  \sigma_0\,Q_{sat}^2\,\frac{2p_T\varepsilon}{(p_T^2+\varepsilon^2)^3}.
 \label{anal}
\end{eqnarray}

We have verified that analytical calculations of referred integrals can be also done if we consider a fixed 'effective' anomalous dimension. Taking the result presented in Eq. ({\ref{anal}), an explicit expression for the hadron cross section is obtained,
\begin{eqnarray}
\frac{d \sigma(pp\to \gamma
X)}{dyd^{2}\vec{p}_{T}}  &\approx &
\frac{\alpha_{em}\sigma_0Q_{sat}^2}{2\pi^2}\int_{x_{1}}^{1}\frac{d\alpha}{\alpha}
 F_{2}^{p}\left(\frac{x_{1}}{\alpha},Q^2\right)\nonumber \\
&\times &\left \{ m_q^2\alpha^4\left[ \frac{p_T^2}{(p_T^2+\varepsilon^2)^4}\right] 
   +   [1+(1-\alpha)^2]\left[ \frac{p_T^4}{2(p_T^2+\varepsilon^2)^4}\right]
\right \} \nonumber
\end{eqnarray}

Therefore, the anomalous dimension for models relying on extended geometric scaling play an important role in the functional form of the $p_T$ spectrum of prompt photon production in hadron-hadron collisions.

As a summary, we showed that direct photon production can be addressed in the NLO QCD approach and in the color dipole picture. In particular, direct photon production can be described within the
color dipole approach without any free parameters by using dipole cross section determined from current  phenomenology in DIS. In central rapidities at RHIC and Tevatron, saturation effects do not play a significant role for direct photon production at the given experimental range of $p_{T}$. This situation can be changed at the LHC even at midrapidities. We also verified that the color dipole formulation including DGLAP evolution (b-SAT model) provides a better description of data at large transverse momentum
compared to the dipole models using fixed or running anomalous dimension. We write down also analytical results relevant for the calculation of $p_T$ spectrum of direct photons. Our results corroborates the statements from Ref. \cite{Kope_gamma} and call attention for the large-$x$ effects that should be added to the corresponding phenomenology for RHIC and Tevatron.

\section*{Acknowledgments}

One of us (MVTM) would like to thank the hospitality of organizers and the lively atmosphere of the Workshop on {\it 3rd Hard Probes Conference 2008}, 8-14 June 2008, Galicia - Spain, where this work was presented.   The authors are grateful to V.P. Gon\c{c}alves, M.A. Betemps and A. Rezaeian for useful information. This work was supported by CNPq, Brazil.

\end{document}